\documentclass[aip,
jcp,
amsmath,
amssymb,
reprint, 
citeautoscript,
]{revtex4-2}

\usepackage{graphicx}

\usepackage[caption=false,labelformat=simple]{subfig}

\usepackage{dcolumn}
\usepackage{hyperref}

\begin{document}

\preprint{}

\title[
Microscopic Mechanisms of
Diffusion Dynamics
: A Comparative Efficiency Study of Event-Chain Monte Carlo Variants in Dense Hard Disk Systems
]{
Microscopic Mechanisms of
Diffusion Dynamics
: A Comparative Efficiency Study of Event-Chain Monte Carlo Variants in Dense Hard Disk Systems
}

\author{Daigo Mugita}
\affiliation{Graduate School of Engineering, Nagoya Institute of Technology,
Nagoya, 466-8555, Japan}

\author{Masaharu Isobe}
\email{isobe@nitech.ac.jp}
\affiliation{Graduate School of Engineering, Nagoya Institute of Technology,
Nagoya, 466-8555, Japan}

\date{\today}

\begin{abstract}
In molecular simulations, efficient methods for investigating equilibration and slow relaxation in dense systems are crucial yet challenging. This study focuses on the diffusional characteristics of monodisperse hard disk systems at equilibrium, comparing novel methodologies of event-chain Monte Carlo variants, specifically the Newtonian event-chain and straight event-chain algorithms. We systematically analyze both event-based and CPU time-based efficiency in liquid and solid phases, aiming to elucidate the microscopic mechanisms underlying structural relaxation. Our results demonstrate how chain length or duration, system size, and phase state influence the efficiency of diffusion dynamics, including hopping motion. This work provides insights into optimizing simulation techniques for highly packed systems and has the potential to improve our understanding of diffusion dynamics even in complex many-body systems.
\end{abstract}


\maketitle

\section{\label{sec:1} Introduction}
In molecular simulations\cite{krauth_2006, allen_2017, frenkel_2023},
the implementation of efficient numerical methods for investigating equilibration and slow relaxation (rare events) in dense systems has been actively investigated and become a source of fascinating and challenging issues. These methods are crucial for studying various physical systems, including supercooled liquids, glass/jamming transitions, protein folding processes, complex-shaped molecular or polymer systems, and others. To achieve sufficient sampling of rare events and obtain statistically accurate and valid data, substantial computational resources are often required, particularly for high-density and large-scale particle systems, even in simple monodisperse systems. In some cases, the computational cost may exceed practical time constraints.

The two main standard methodologies for molecular simulation in many-body systems are molecular dynamics (MD)\cite{rapaport_2004}, which uses Newton's equations of motion, and Markov chain Monte Carlo (MCMC)\cite{metropolis_1953}. Both methods equilibrate many-body systems, leading to equilibrium states, as established in statistical mechanics\cite{krauth_2006}. However, the paths and efficiency to 
achieve equilibration and equilibrium (diffusion) dynamics
can differ significantly between these methods.
In
the history of molecular simulation, researchers have actively studied various novel algorithms focusing on equilibration efficiency or relaxation time of equilibrium 
states (e.g., correlation functions).
These studies serve as benchmarks for improving simulation techniques.

In hard disk/sphere systems, a seminal paper on computer simulation,
published in 1957, provided groundbreaking physical insights and methodologies for understanding
the solid-fluid phase transition\cite{alder_1957, wood_1957}. This pioneering study employed both MCMC and
the event-driven molecular dynamics (EDMD) method\cite{alder_1959, isobe_1999, rapaport_2004}, which 
subsequently led to the discovery of
the ``Alder transition''\cite{hiwatari_2009, isobe_2016a, li_2022}.
In the melting problem of hard disk systems\cite{alder_1962, bernard_2011, engel_2013, kosterlitz_1973, halperin_1978, nelson_1979, young_1979, strandburg_1988}, 
simulations
on a large scale was essential to elucidate the controversial nature of the phase transition. Insufficient equilibration caused different conclusions and controversy for many decades.

A significant milestone in improving equilibration efficiency is the event-chain Monte Carlo (ECMC)
method\cite{bernard_2009, krauth_2021}.
ECMC is an efficient sampling 
technique
that incorporates the concepts of factorization and lifting into MCMC\cite{tartero_2024}.
In the most basic reversible MCMC with a Metropolis filter, the rejection of proposed transitions is determined by changes in the system's total potential. However, 
ECMC employs a different approach:
the potential is factorized, and transitions are evaluated uniquely by individual factors\cite{michel_2014}. When a transition is rejected, the system state is irreversibly altered using the lifting technique\cite{kapfer_2017}. 
Through this mechanism,
ECMC achieves fast sampling while maintaining global balance rather than detailed balance.

In the case of hard disk/sphere systems, ECMC 
consists of successive displacements of the active particles through
a sequence of particle collisions (event-chain).
Over
the past decade, researchers have actively developed several variants of ECMC\cite{krauth_2021, klement_2019, hollmer_2022a}, while also thoroughly investigating benchmarks such as the relaxation time of equilibrium correlation functions across different methods\cite{bernard_2009, engel_2013, isobe_2015, klement_2019}.
A
recent 
notable
advancement is the Newtonian event-chain (NEC) algorithm\cite{klement_2019}, 
which incorporates collision rules for velocities to determine the direction of particle
displacement, analogous
to EDMD. NEC has demonstrated 
significant
advantages in efficiency for calculating
melting processes, nucleation rates, and diffusion coefficients. Moreover, it
has been 
successfully applied
to simulate hard anisotropic particles without approximations, incorporating an implementation of efficient contact detection\cite{klement_2021}.

In general, the relaxation
of particle positions, i.e., structural relaxation, is 
exceptionally challenging, particularly
in highly dense systems, 
due to
the dominant excluded-volume effect. 
Structural relaxation, characterized by particle displacement from the origin, is of primary interest in studies of slow dynamics in glassy physics\cite{chandler_2010, berthier_2011, biroli_2013, royall_2015, torquato_2010, scalliet_2022}. In these investigations, mean square displacement, intermediate scattering function, and non-Gaussian parameter are often key metrics of correlation functions used to quantify structural relaxation.

Numerical approaches to structural relaxation and particle diffusion in EDMD and ECMC differ significantly, making it challenging to quantify their respective capabilities.
In EDMD, even when particles and clusters composed of dozens of particles 
become
trapped in metastable states\cite{isobe_2012b}, the velocities can attain the Maxwell-Boltzmann distribution after just a few collisions. In 
contrast, ECMC drives systems through
sequential collisions of particles 
forming a chain-like structure, focusing solely on the displacement of particle
positions (diffusion). The optimal performance of these algorithms strongly depends on
various
parameters, including
chain length, physical properties of the system, and system size. 
Several studies have previously reported
comparisons of efficiency 
between methodologies\cite{engel_2013, isobe_2016a, isobe_2015, klement_2019}. However, the microscopic mechanisms underlying these dependencies, especially 
regarding diffusional characteristics in high-density states, remain elusive.

To elucidate the 
crucial
factors
influencing efficiency in structural relaxation and the microscopic mechanisms of 
diffusion
dynamics across different methods, we systematically 
focus on
the diffusional characteristics in terms of 
both
event-based and CPU time-based efficiency. Our study compares the
NEC
algorithm 
with
the straight event-chain (SEC) algorithm, 
the latter being
the simplest variant of ECMC. 
We conduct this investigation in a simple monodisperse hard disk system at equilibrium, examining both liquid and solid phases.

This paper is organized as follows. In Sec.~\ref{sec:2}, we describe our model and the numerical settings, including the definitions of diffusion coefficient and the related physical properties. Section~\ref{sec:3} presents results that compare efficiency by changing the chain length and size of the system for both the liquid and the solid phase. Concluding remarks and discussion are summarized in Sec.~\ref{sec:4}.

\section{\label{sec:2} Model and simulation methods}
We focus on two-dimensional (2D) monodisperse systems consisting of $N$ hard disks without any external force, friction, or dissipation between disks. The disks of radius $\sigma$ are placed in a $L_x \times L_y (= A)$ rectangular box (i.e., $L_y/L_x = \sqrt{3}/2$) with periodic boundary conditions imposed in both directions. The basic units of the system are set by the mass of one disk $m$, the diameter of the disk $d$ ($= 2\sigma$), and the energy $1/\beta$, from which we derive the unit of time as $d\sqrt{\beta m}$, where $\beta$ is $1/k_{\mathrm{B}}T$ using the Boltzmann constant $k_{\mathrm{B}}$ and the temperature $T$.

Hard sphere systems in 2D and 3D are governed by solid-fluid phase transitions (the so-called ``Alder transition''), with the packing fraction $\nu = N\pi\sigma^2/A$ (2D) serving as the primary control parameter. Recent large-scale hard disk simulations have revealed a series of phase changes: from liquid ($\nu < 0.700$), through a coexistence region ($0.700 < \nu < 0.716$) and a hexatic phase ($0.716 < \nu < 0.720$), to solid (crystal) ($\nu > 0.720$)\cite{alder_1962, bernard_2011, engel_2013}. For our simulations, we selected packing fractions of $\nu = 0.450$ for the liquid phase and $\nu \geq 0.720$ for the solid phase. Preliminary results for the liquid phase have been reported elsewhere\cite{banno_2022}.

As initial states, we carefully prepared equilibrated configurations for each packing fraction through long-time equilibration using EDMD runs.
Production runs were carried out using EDMD and three different ECMC sampling algorithms. In ECMC, trial moves are performed through a chain of particle collisions (event-chain). The event-chain is terminated when the total displacement of the particles reaches a certain threshold, known as the chain length $L_{\mathrm{c}}$. At this point, a particle is randomly selected, and a new event-chain is initiated from that particle. We employ three ECMC variants:
\begin{enumerate}
\item Straight event-chain algorithm alternating $x$- and $y$-axes directions (SEC-xy)
\item Straight event-chain algorithm with uniformly random directions (SEC-all)
\item Newtonian event-chain algorithm (NEC)
\end{enumerate}
In the SEC algorithms, the direction of particle movement is restricted within one event-chain. SEC-xy alternates the movement direction along the $x$- and $y$-axis for each event-chain, while SEC-all uniformly selects the movement direction from all possible directions. The NEC variant introduces Maxwell-Boltzmann velocity distributions of disks, similar to the EDMD, to determine the direction of displacement based on collision rules of velocities. In NEC, we use the chain duration of accumulated time $T_{\mathrm{c}}$ rather than the chain length $L_{\mathrm{c}}$, as each disk's velocity affects its displacement length by multiplication by the root mean square velocity $v_{\mathrm{rms}}=\sqrt{\left< v^2\right>}$, where $\langle v^2 \rangle$ is the ensemble average of the squared velocity.

We calculate the diffusion coefficients, 
$D_{\mathrm{ev}}$
and $D_{\mathrm{cpu}}$, based on the number of 
events $N_{\mathrm{ev}}$
and CPU time $t_{\mathrm{cpu}}$, respectively\cite{klement_2019, banno_2022}. Each coefficient was derived from the mean square displacement (MSD) $\langle|\Delta \mathbf{r} - \langle \Delta \mathbf{r} \rangle|^2\rangle = \sum_{i=1}^N |\Delta \mathbf{r}_i - \langle \Delta \mathbf{r} \rangle|^2 / N$, using the following equations in two dimensions:
\begin{equation}
D_{\mathrm{ev}}
= \lim_{
N^*_{\mathrm{ev}}
 \to \infty} \frac{\langle|\Delta \mathbf{r} - \langle \Delta \mathbf{r} \rangle|^2\rangle}{4 
N^*_{\mathrm{ev}}
},
\label{eq:D}
\end{equation}
\begin{equation}
D_{\mathrm{cpu}} = \lim_{t_{\mathrm{cpu}} \to \infty} \frac{\langle|\Delta \mathbf{r} - \langle \Delta \mathbf{r} \rangle|^2\rangle}{4 t_{\mathrm{cpu}}}.
\label{eq:Dcpu}
\end{equation}
Here, $\Delta \mathbf{r}_i$ is the displacement vector of the particle $i$ after
$N_{\mathrm{ev}}$
times 
events, $\langle \Delta \mathbf{r} \rangle$ is the mean displacement vector averaged over all particles in the system, and 
$N^*_{\mathrm{ev}}$
is the number of displacements per particle ($=N_{\mathrm{ev}}/N$).
In this study, with a focus on the efficiency of structural relaxation, we subtracted the displacement by $\langle \Delta \mathbf{r} \rangle$ to eliminate the flow component.

All simulations were performed on a single core of an Intel Xeon E5-1660, 3.3 GHz. To optimize the codes for all algorithms, we implemented a grid mapping technique (the exclusive grid particle method)\cite{isobe_1999} for efficient management of each particle's neighbor list.

\section{\label{sec:3} Results}
\subsection{Liquid Phase}
\subsubsection{Mean square displacement}

\begin{figure}
    \centering
    \includegraphics[scale=0.13]{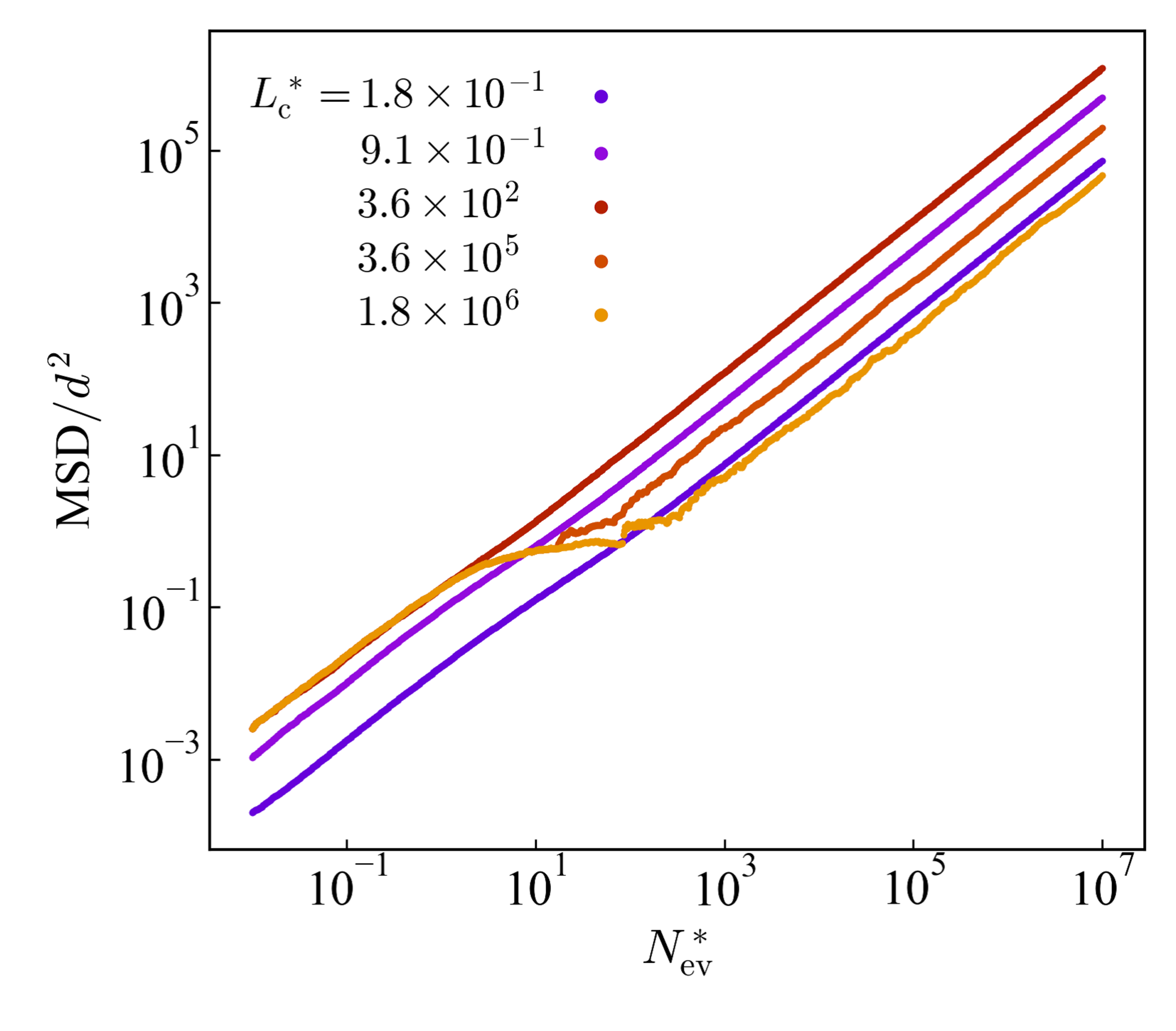}
    \caption{Evolution of the MSD as a function of 
$N_{\mathrm{ev}}^*$
for the SEC-all with various chain lengths in the liquid state at $(N, \nu) = (65536, 0.450)$. $L_{\mathrm{c}}^*$ is defined as $L_{\mathrm{c}}/d$, where $d$ is the disk diameter.}
    \label{fig:MSD_liquid}
\end{figure}

Figure~\ref{fig:MSD_liquid} illustrates the evolution of the mean square displacement (MSD) as a function of 
$N_{\mathrm{ev}}^*$
for the SEC-all with various chain lengths in the liquid state at $(N, \nu) = (65536, 0.450)$. Here, $L_{\mathrm{c}}^*$ is defined as $L_{\mathrm{c}}/d$, where $d$ is the disk diameter. For relatively long chain lengths, a plateau region is observed in the intermediate stage. 
This is due to unidirectional displacement, similar to domain flows, being organized within a single event-chain. This organization decreases diffusion relative to neighboring particles, causing an overall reduction in diffusion. This mechanism will be addressed in detail later in Sec.~\ref{sec:3-1-3}.

Among all methods used in this study, SEC-all requires the highest number of displacements to reach the regime of normal diffusion. Normal diffusive behavior in SEC-all was observed at
$N_{\mathrm{ev}}^*\ge 10^6$, even in high-density regions
where the packing fraction $\nu \le 0.740$,
and for 
long
chain lengths (see also
the results of NEC in the solid phase,
Fig.~\ref{fig:MSD_D_solid}(a)).
Consequently, MSD values
for $10^6 \le N_{\mathrm{ev}}^* (\le 10^7)$
were deemed relevant for investigating diffusion characteristics. Hereafter, we use this threshold value
$N_{\mathrm{ev}}^*= 10^6$ for all subsequent analyses throughout this paper.
For the calculation of $D_{\mathrm{cpu}}$, we used a range of $t_{\mathrm{cpu}}$ that corresponds to this $N_{\mathrm{ev}}^*$ range.
The validity and limitations of this $N_{\mathrm{ev}}^*$ range for MSD for the diffusion coefficient calculation in the high-density region will be discussed in Sec.~\ref{sec:3-2-2}.

\subsubsection{Chain length or duration dependence of diffusion coefficients}
\begin{figure*}
    \centering
    \includegraphics[scale=0.12]{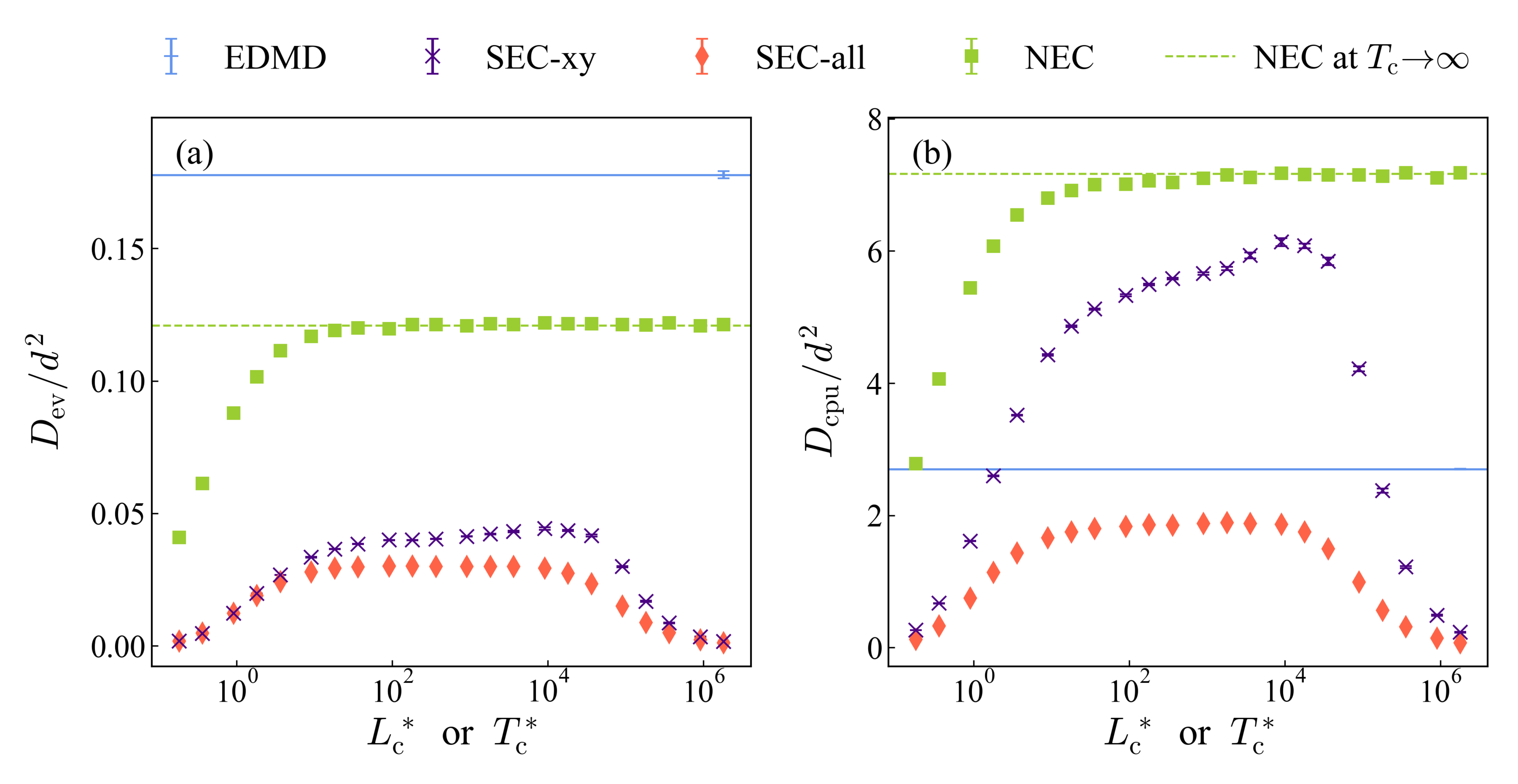}
    \caption{Chain length 
or duration
dependence of (a) 
$D_{\mathrm{ev}}$
and (b) $D_{\mathrm{cpu}}$ for EDMD, 
SEC-xy, SEC-all, and NEC
in the liquid state at $(N, \nu) = (65536, 0.450)$, averaged over $25$ simulations.
$T_{\mathrm{c}}^*$ is defined as $T_{\mathrm{c}}v_{\mathrm{rms}}/d$.
}
    \label{fig:DDcpu_liquid}
\end{figure*}

The diffusion coefficients of ECMC exhibit a three-stage behavior with respect to chain length
or duration (Fig.~\ref{fig:DDcpu_liquid}):
\begin{enumerate}
\item Initial Stage: As the chain length
or duration
 increases, the diffusion efficiency improves for all ECMC variants.
\item Intermediate Stage: NEC and SEC-all show a plateau in efficiency, whereas SEC-xy exhibits a slight increase.
\item Final Stage: Both SEC variants experience a decrease in efficiency, whereas NEC maintains its highest efficiency; in NEC, even as $T_{\mathrm{c}} \to \infty$ (i.e., without resampling), efficiency appears not to decrease.
\end{enumerate}

These trends for both SEC variants are fairly consistent with those reported previously in Ref.~\onlinecite{bernard_2009}. The efficiency, ranked from highest to lowest in terms of
event, is EDMD, NEC, SEC-xy, and SEC-all, which is consistent with the results reported for 3D systems in Ref.~\onlinecite{klement_2019}. Conversely, the order of efficiency from highest to lowest in terms of CPU time (which is crucial from a practical performance perspective), considering the results at the most efficient chain length
or duration, is NEC, SEC-xy, EDMD, and SEC-all.
However, for $N \le 16384$, the most efficient $D_{\mathrm{cpu}}$ for SEC-xy surpassed that of NEC, resulting in the order at the most efficient chain length
or duration: SEC-xy, NEC, EDMD, and SEC-all.

Table~\ref{tab:liquid} 
summarizes the efficiencies of diffusion coefficients in terms of event and CPU time relative to the values of SEC-all. In systems with small sizes or high densities, we observed relatively large fluctuations in efficiency, which varied depending on chain length or duration between independent runs with identical conditions ($N$ and $\nu$). To provide a robust measurement, we selected the top five samples of $D_{\mathrm{ev}}$ and $D_{\mathrm{cpu}}$ from the ensemble data obtained for the diffusion coefficient by each method for fixed $N$ and $\nu$. These samples were averaged to produce $D_{\mathrm{ev}}^{\mathrm{top5}}$ and $D_{\mathrm{cpu}}^{\mathrm{top5}}$, respectively. We consistently employed these averaged values to evaluate efficiency throughout our study, ensuring a reliable comparison. These values were derived by ranking the diffusion coefficients in descending order from the ensemble data for each corresponding chain length or duration.
Taking into account the trends in the efficiency
in terms of CPU time
relative to
that in terms of event, the efficiency of NEC remained relatively unchanged, while the efficiency of EDMD decreased drastically, and the efficiency of SEC-xy increased. In EDMD, since all particles are involved and updated in each event, the efficiency in terms of 
event is relatively high. However, the efficiency in terms of CPU time generally decreases because of the need to calculate the minimum collision times for all particles in each event. The substantial improvement in the efficiency of SEC-xy in terms of CPU time can be primarily attributed to the reduction in the number of neighboring particles surveyed as potential collision candidates using grid mapping techniques. This reduction in computational cost is possible because the direction of displacement is restricted to only two axes, resulting in an optimization of the number of displacement distance evaluations when the grid lattice aligns with the $x$- and $y$-axes.

\begin{table}[!t]
\caption{
Comparison of diffusion efficiencies across different methods in liquid phase at $(N, \nu) = (65536, 0.450)$, estimated in terms of events and CPU time. Those are normalized relative to the SEC-all method.
}
\label{tab:liquid}
\begin{ruledtabular}
\begin{tabular}{lcccc}
& SEC-all & SEC-xy & NEC & EDMD \\ \hline
$D_{\mathrm{ev}}^{\mathrm{top5}}$ or $D_{\mathrm{ev}}$ & $1.00$ & $1.43$ & $4.06$ & $5.93$ \\
$D_{\mathrm{cpu}}^{\mathrm{top5}}$ or $D_{\mathrm{cpu}}$ & $1.00$ & $3.18$ & $3.83$ & $1.44$
\end{tabular}
\end{ruledtabular}
\end{table}

\subsubsection{System size dependence}
\label{sec:3-1-3}
\begin{figure*}
    \centering
    \includegraphics[scale=0.19]{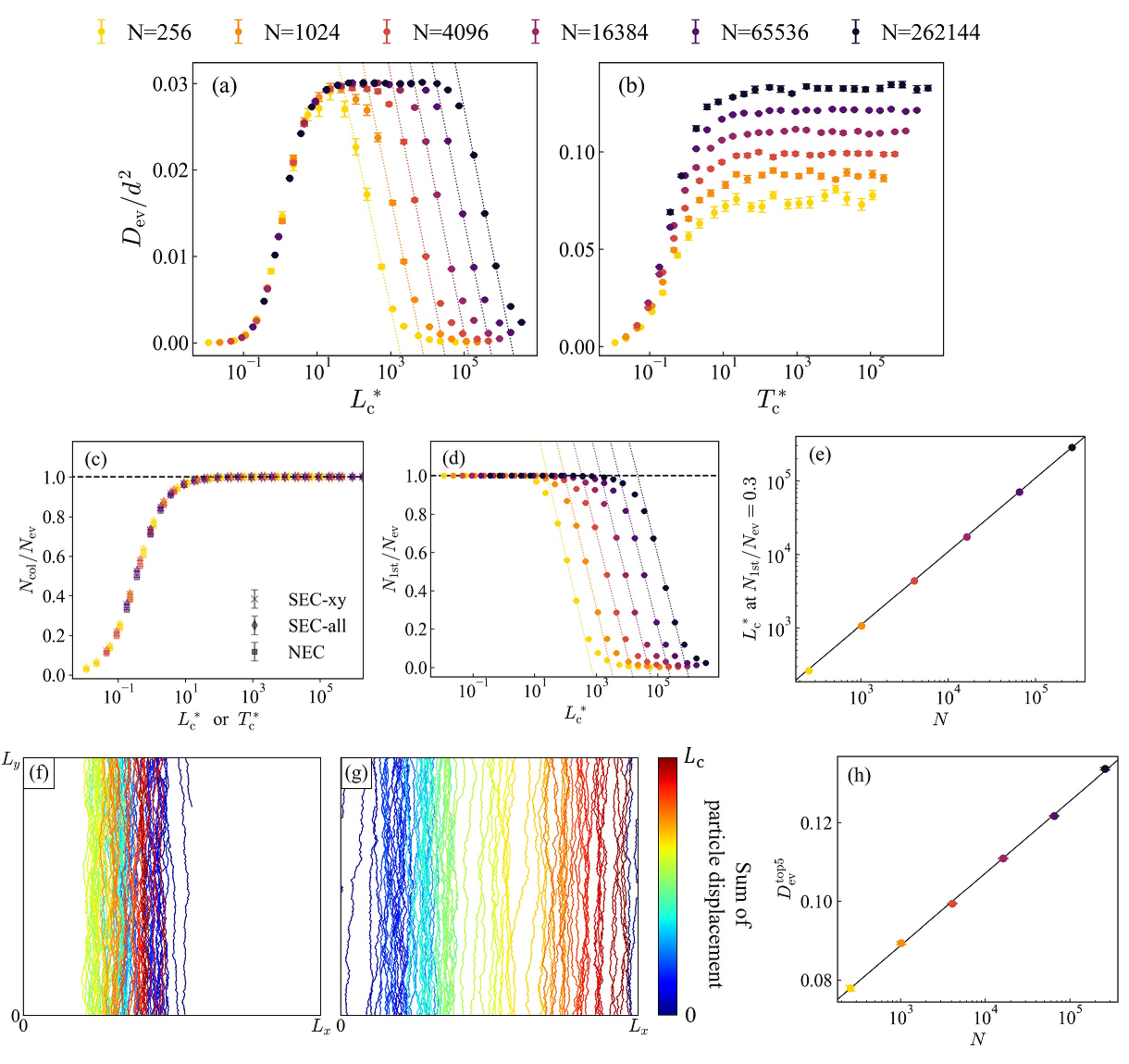}
    \caption{    
The system size dependence of 
$D_{\mathrm{ev}}$
for a wide range of chain lengths or durations for (a) SEC-all and (b) NEC in a liquid state at $\nu = 0.450$. 
For $N=262144$, data were averaged over five simulations, while the others were averaged over 25 simulations. (c) Chain length or duration and system size dependencies of $N_{\mathrm{col}}$ per event in simulations up to $N_{\mathrm{ev}}^*=6.0\times10^6$ by SEC-xy, SEC-all, and NEC for $N=256$, $4096$, and $65536$.
The coloring corresponds to the system size 
same as
in (a) and (b).
(d) Chain length and system size dependencies of $N_{\mathrm{1st}}$ per event in simulations up to $N_{\mathrm{disp}}^*=10^3$ by SEC-all.
(e) The system size dependence of the chain length 
in SEC-all at which the decreasing curves fitted with a logarithmic function (dotted lines in (d)) crosses the line of $N_{\mathrm{1st}}/N_{\mathrm{ev}}=0.3$.
(f)(g)
Two typical successive active particles' trajectories within a single event-chain with the length $L_{\mathrm{c}}^*=9.0\times 10^3$ in the direction along the $y$-axis of SEC-xy in a liquid state at $(N, \nu) = (65536, 0.450)$, where the color gradient of the trajectories corresponds to the accumulated displacement of the active particles. The direction of drift aligns to the $x$-axis and shifts multiple times in (f), while it persists until the end of the event-chain in (g).
(h)
The system size dependence of the 
$D_{\mathrm{ev}}^{\mathrm{top5}}$
by NEC.
} 
    \label{fig:system_size}
\end{figure*}

To investigate the mechanism underlying the three-stage behavior with respect to chain length
or duration, we examine the system size dependence of 
$D_{\mathrm{ev}}$. Figure~\ref{fig:system_size}(a) illustrates the system size dependence of SEC-all. Although no significant differences in efficiency were observed during the initial stage between system sizes, notable variations emerged in the chain length at which the final stage begins, immediately following the intermediate stage.
Similar system size dependency was observed in SEC-xy as well.
Figure~\ref{fig:system_size}(b) represents the system size dependence of NEC, where the diffusion efficiency increases with system size.

Figure~\ref{fig:system_size}(c) presents the dependence of the number of collisions $N_{\mathrm{col}}$ per
event
on chain length
or duration. For shorter chain lengths
or durations
in ECMC, the accumulated displacement distance exceeds the chain length 
or duration
before the active disk collides with others, resulting in displacements without collision events. This behavior is more akin to acceptance trials in local MCMC than to event-chain events (i.e., collisions). As the chain length 
or duration
increases, the ratio of collisions to
events
approaches $N_{\mathrm{col}}/N_{\mathrm{ev}}=1$ at $L_{\mathrm{c}}^*$ or $T_{\mathrm{c}}^* \sim10^1$, indicating that collisions within the event-chains become dominant.

We found that the initial stage of 
$D_{\mathrm{ev}}$
also ends at $L_{\mathrm{c}}^*$ or $T_{\mathrm{c}}^* \sim 10^1$, and the increasing trend of $N_{\mathrm{col}}$ aligns with the behavior of the diffusion coefficient for both SEC-all and NEC up to the intermediate stage. This suggests that the initial stage increase in the diffusion coefficient arises from the ratio of
events
without collisions to those with collisions, representing a transition from MCMC-like to ECMC-like behavior. 
To confirm the dominant
factors
in the initial stage, we examined the $N_{\mathrm{col}}/N_{\mathrm{ev}}$
curves for various system sizes
at $N=256$, $4096$, and $65536$, which show universal behavior as illustrated in Fig.~\ref{fig:system_size}(c). In the intermediate stage, this explanation does not fully account for the slight increase in SEC-xy, suggesting the existence of another undiscovered microscopic mechanism.

Figure~\ref{fig:system_size}(d) illustrates the chain length dependence of $N_{\mathrm{1st}}$, which represents the number of times a particle becomes active for the first time within an event-chain, normalized per event for SEC-all. In SEC algorithms, for short chain lengths, a particle typically becomes active only once within a single event-chain, resulting in $N_{\mathrm{1st}}/N_{\mathrm{ev}}\sim1$. As the chain length increases and the event-chain traverses periodic boundaries, it encounters particles that have previously been activated, causing $N_{\mathrm{1st}}$ to decrease. For very long chain lengths, as demonstrated in Fig.~\ref{fig:system_size}(d), $N_{\mathrm{1st}}$ approaches zero, indicating that in the majority of events, the displacement occurs primarily in particles that have already experienced activation within the same event-chain.

The final stage in Fig.~\ref{fig:system_size}(a) and the decreasing part in Fig.~\ref{fig:system_size}(d) exhibit remarkably similar trends, both demonstrating logarithmic decay. Figure~\ref{fig:system_size}(e) illustrates the system size dependence of the chain length at which the function, fitted with a logarithmic function of $N_{\mathrm{1st}}/N_{\mathrm{ev}}$ for SEC-all (depicted as dotted lines in Fig.~\ref{fig:system_size}(d)), decreases to $0.3$. Our analysis reveals that the chain length at which $N_{\mathrm{1st}}$ begins to decrease is proportional to the system size. Independently, an alternative analysis for $D_{\mathrm{ev}}$ similarly demonstrates proportionality to the system size, indicating a strong correlation between these two properties.

To investigate why 
$N_{\mathrm{1st}}$ 
decreases proportionally
to 
system size $N$ $(\sim L_x^2$ or $L_y^2)$ rather than 
system length $L_x$ (or $L_y$) in the SEC algorithm, we examined 
successive activated particle trajectories within a single event-chain. We used a displacement length $L_{\mathrm{c}}^*=9.0\times 10^3$ for a liquid phase at $(N, \nu) = (65536, 0.450)$ in SEC-xy, as shown in Figs.~\ref{fig:system_size}(f) and (g). The trajectory color gradient corresponds to the active particles' accumulated displacement.
In SEC-xy, successive collision positions do not follow a strictly straight line but exhibit drift with slight fluctuations in directions different from particle displacement. These drifts, observed even in the liquid phase, are unlikely driven by crystal structure orientational order. This drift results in a slightly different chain position when traversing the system under periodic boundary conditions, effectively sampling in two dimensions rather than one.
In Fig.~\ref{fig:system_size}(f), the drift direction aligns with the $x$-axis and shifts multiple times, while in (g), it persists until the event-chain's end, spreading sampling regions across the entire system. We confirmed a similar phenomenon in SEC-all. This observation explains the linear increase of $L_{\mathrm{c}}$ with $N$ observed in Fig.~\ref{fig:system_size}(e).

Figs.~\ref{fig:system_size}(f) 
suggest
a mechanism for 
decreased
diffusion efficiency at long chain lengths. 
The chain length 
$L_{\mathrm{c}}^*=9.0\times 10^3$ corresponds 
to both the point where the event-chain width potentially spans the entire system and where $D_{\mathrm{ev}}$ begins to decrease.
In principle, 
event-chain activated particles contribute to diffusion by separating their relative positions from
neighbors. 
However,
long event-chains like those in Figs.~\ref{fig:system_size}(f) and (g) produce
unidirectional flow with neighbors
in the same direction.
Since the diffusion coefficient is defined by Eq.~(\ref{eq:D}), which includes $\langle \Delta \mathbf{r} \rangle$, 
it
decreases under 
unidirectional flow. 
Note that multiple shifts in drift direction, as in Fig.~\ref{fig:system_size}(f), cause domain-like flow in partial system regions, which reduces the diffusion coefficient within the flow domain.

Based on above considerations, the behavior of MSD in Fig.~\ref{fig:MSD_liquid} can be interpreted as follows: In the case of long chain lengths in SEC-all, during the initial stage, the MSD increases linearly with respect to $N_{\mathrm{ev}}^*$. Eventually, around $N_{\mathrm{ev}}^* = 10^1$, where the number of events corresponds to the particle number at which a particle is activated (displaced) once on average, we observe the emergence of a plateau in Fig.~\ref{fig:MSD_liquid}. This plateau is presumed to be caused by the unidirectional displacement of all particles (flow of the entire system).

In NEC, due to the randomness of the velocity distributed according to the Maxwell-Boltzmann distribution obeying the collision rule of hard disks, unidirectional flow of the entire system as observed in SEC does not occur even with long chain durations, resulting in $\langle \Delta \mathbf{r} \rangle \sim 0$.
As a result, the diffusion coefficient did not decrease even as $T_{\mathrm{c}}^* \to \infty$ in the system used in this study. However, it has been proven that without resampling, the system is not always irreducible\cite{hollmer_2022b}. 
Additionally, in specific sparse hard disk packings, it is known that ``gridlocks'' frequently occur, resulting in limit cycles (collapse or loops) of displacements composed of a few specific particles\cite{hollmer_2022a}.

For NEC, calculating the average 
$D_{\mathrm{ev}}^{\mathrm{top5}}$
of the top five highest 
$D_{\mathrm{ev}}$
values for each system size revealed a logarithmic increase with system size (Fig.~\ref{fig:system_size}(h)). This suggests that the diffusion coefficient of NEC would diverge in the thermodynamic limit if the trend does not change. This curious behavior warrants further investigation.

\subsection{Solid Phase}
While random configurations in the liquid phase and ordered structures in the solid (crystal) phase would exhibit differences in diffusional efficiency as a result of positional and orientational order, investigating abnormal diffusion in densely packed crystalline states presents unique challenges. In highly packed systems, even those composed of monodisperse particles, diffusional characteristics are predominantly governed by hopping motions, short-timescale diffusion between basins in the potential (or entropic) landscape. This subsection systematically examines the efficiency of diffusion in the crystalline state at high packing fractions above the phase transition point ($\nu \ge 0.720$), comparing ECMC variants.

\subsubsection{Mean square displacement and diffusion coefficients}
\begin{figure*}
    \centering
    \includegraphics[scale=0.13]{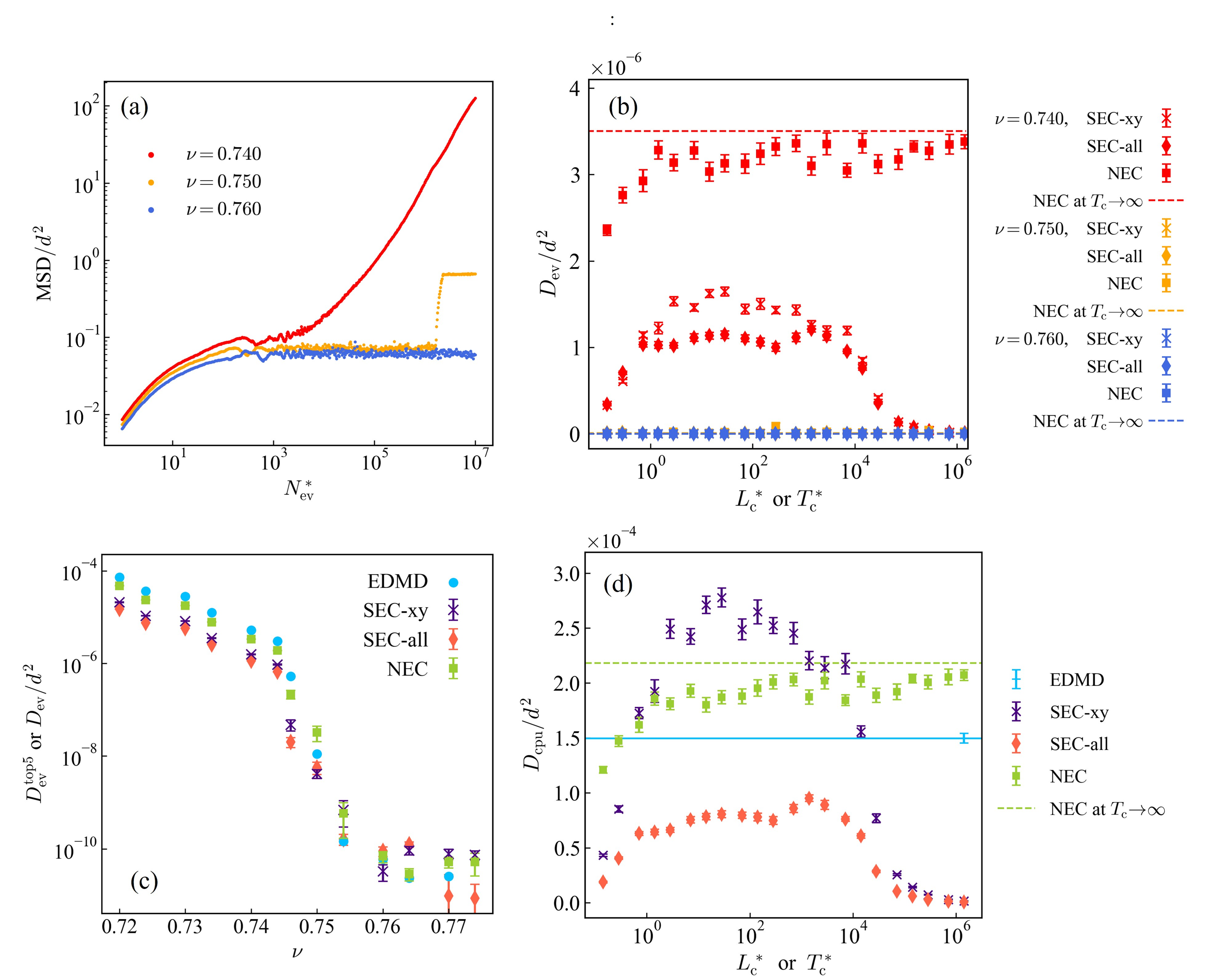}
    \caption{(a) The MSD obtained by NEC at $T_{\mathrm{c}}^*=1.4\times10^2$ in the solid state at packing fractions $\nu=0.740, 0.750,$ and $0.760$ for a system size of $N = 65536$, starting from a specific equilibrium configuration. 
(b) Chain length 
or duration
dependence of 
$D_{\mathrm{ev}}$
for various ECMC variants at packing fractions $\nu=0.740$, $0.750$, and $0.760$ for $N = 65536$, averaged over $25$ independent simulations. (c) Packing fraction dependence of
$D_{\mathrm{ev}}^{\mathrm{top5}}$
in the solid phase for $N=65536$. For EDMD, 
$D_{\mathrm{ev}}$
is displayed instead of 
$D_{\mathrm{ev}}^{\mathrm{top5}}$
due to its independence from chain length
or duration. (d) $D_{\mathrm{cpu}}$ as a function of chain length 
or duration
for EDMD and each ECMC variant at $(N, \nu) = (65536, 0.740)$, averaged over $25$ independent simulations.
    }
    \label{fig:MSD_D_solid}
\end{figure*}

Figures~\ref{fig:MSD_D_solid}(a) and \ref{fig:MSD_D_solid}(b) illustrate the MSD obtained by NEC and the chain length 
or duration
dependence of 
$D_{\mathrm{ev}}$
for ECMC variants at packing fractions $\nu=0.740, 0.750,$ and $0.760$, respectively.
At $\nu=0.740$, 
a plateau region is observed in the intermediate stage, reminiscent of the hallmark two-step relaxation in glassy dynamics.
This behavior is characterized by an initial rapid increase ($\beta$-relaxation), followed by a plateau corresponding to the cage effect. The subsequent $\alpha$-relaxation manifests itself as a transition from this plateau to normal diffusive behavior, characterized by a linear increase of the MSD with time, as also reported in Ref.~\onlinecite{klement_2019}.
The
MSD for NEC increases linearly after the threshold
$N_{\mathrm{ev}}^*\sim 10^6$.
$D_{\mathrm{ev}}$
at $\nu=0.740$
exhibits three-stages of behavior as the chain length 
or duration
increases, similar to the results observed in the liquid phase.

In contrast, at $\nu=0.750$ and $0.760$, there is little to no increase in MSD, and the 
$D_{\mathrm{ev}}$
values remain close to zero for all chain lengths
or durations, indicating an apparent absence of diffusion. However, at $\nu=0.750$, a step-like sudden increase in MSD was 
sometimes
observed within a short interval, as shown in 
Fig.~\ref{fig:MSD_D_solid}(a). This instantaneous large displacement was also observed in all other methods, including EDMD, suggesting the occurrence of hopping motion.
In this study, all calculations of the diffusion coefficients, including those for $\nu > 0.740$, were evaluated based on the same criterion within the range of $10^6 < N_{\mathrm{ev}}^* < 10^7$. The validity of this range will be addressed in a subsequent section.

Figure~\ref{fig:MSD_D_solid}(c) shows the dependence of 
$D_{\mathrm{ev}}^{\mathrm{top5}}$
on packing fraction, where 
$D_{\mathrm{ev}}^{\mathrm{top5}}$
is the average of the five highest 
$D_{\mathrm{ev}}$
values for each packing fraction. Note that since EDMD is independent of chain length
or duration, it is represented by 
$D_{\mathrm{ev}}$
instead of 
$D_{\mathrm{ev}}^{\mathrm{top5}}$. At $\nu \le$
$0.744$, as the packing fraction increases, the values of 
$D_{\mathrm{ev}}^{\mathrm{top5}}$
gradually decrease while maintaining the efficiency ranking order: EDMD, NEC, SEC-xy, and SEC-all. However, at $\nu \ge 0.746$, the efficiency ranking becomes unstable and fluctuates.
(At $\nu= 0.746$, the relative ranking of efficiencies among methods remains unchanged. However, the rate at which efficiency decreases with respect to $\nu$, as well as the magnitude of efficiency differences between methods, differs from those observed at lower $\nu$ values.)
At $\nu \ge 0.760$, $D_{\mathrm{ev}}^{\mathrm{top5}}$ assumes very low values. These results indicate that stable diffusion can be observed at $\nu \le0.744$
in systems with $N = 65536$.

Figure~\ref{fig:MSD_D_solid}(d) shows the chain length 
or duration
dependence of practical diffusion efficiency in terms of CPU time in the solid phase at $\nu=0.740$. Similar to the liquid phase, the efficiency 
in terms of CPU time
relative to 
that in terms of event
based on the results of SEC-all, decreased for EDMD and increased for SEC-xy
(see also Table~\ref{tab:solid}). Notably, SEC-xy with $L_{\mathrm{c}}^*$ in the range of $10^1$ to $10^3$ demonstrates the highest performance, surpassing the CPU time efficiency of NEC.

\begin{table}[!t]
\caption{
Comparison of diffusion efficiencies across different methods in solid phase at $(N, \nu) = (65536, 0.740)$, estimated in terms of events and CPU time. Those are normalized relative to the SEC-all method.
}
\label{tab:solid}
\begin{ruledtabular}
\begin{tabular}{lcccc}
& SEC-all & SEC-xy & NEC & EDMD \\ \hline
$D_{\mathrm{ev}}^{\mathrm{top5}}$ or $D_{\mathrm{ev}}$ & $1.00$ & $1.35$ & $2.93$ & $4.54$ \\
$D_{\mathrm{cpu}}^{\mathrm{top5}}$ or $D_{\mathrm{cpu}}$ & $1.00$ & $3.06$ & $2.38$ & $1.74$
\end{tabular}
\end{ruledtabular}
\end{table}

\subsubsection{Microscopic mechanisms of diffusion}
\label{sec:3-2-2}
\begin{figure*}
    \centering
    \includegraphics[scale=0.10]{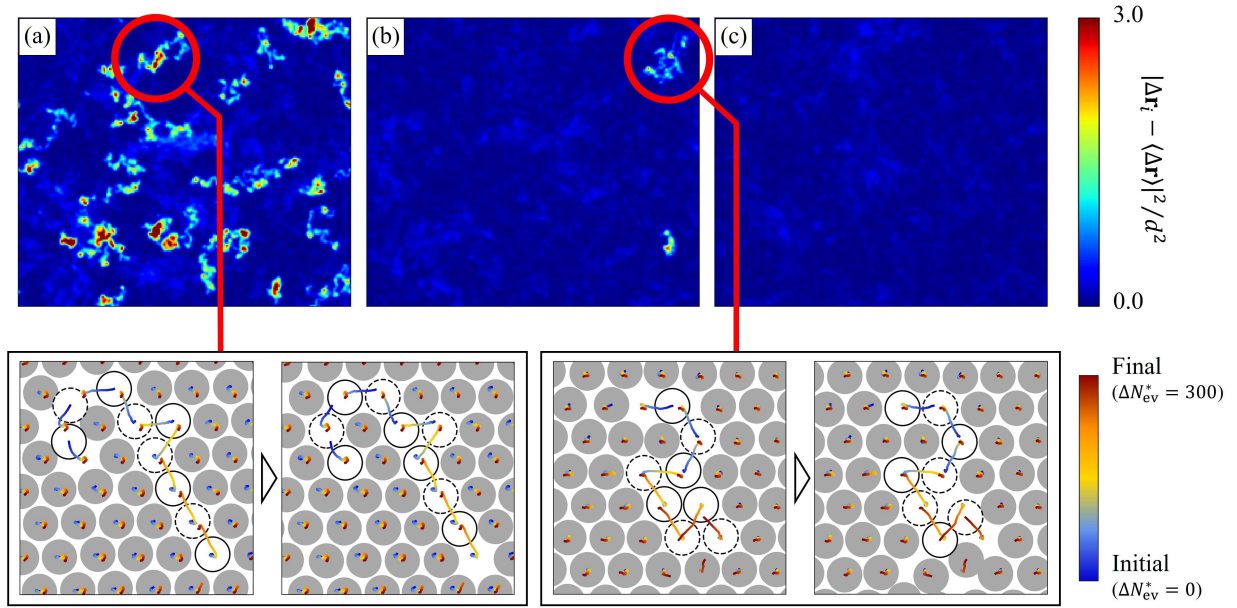}
    \caption{(Top) Squared displacement field $|\Delta \mathbf{r}_i - \langle \Delta \mathbf{r} \rangle|^2/d^2$
for 
$N_{\mathrm{ev}}^*=10^4$ obtained by NEC at 
$T_{\mathrm{c}}^*=28$
for a system size of $N=65536$ at (a) $\nu=0.740$, (b) $0.750,$ and (c) $0.760$. The fields were coarse-grained and smoothed using a Gaussian filter for clear visualization, dividing the system into a $256 \times 222$ grid. (Bottom) Typical successive hopping motion of particle displacements within 
$N_{\mathrm{ev}}^*=300$ at 
(Left)
$\nu=0.740$ and 
(Right)
$0.750$.
The particle trajectories are coarse-grained.
}
    \label{fig:hopping}
\end{figure*}

To elucidate the mechanism underlying the packing fraction dependence of diffusion in the solid phase, we investigated the squared displacement field $|\Delta \mathbf{r}_i - \langle \Delta \mathbf{r} \rangle|^2$ 
for $N_{\mathrm{ev}}^*=10^4$
using NEC, as illustrated in Fig.\ref{fig:hopping}(top). The results revealed dynamic heterogeneity in the system at $\nu=0.740$
and $\nu=0.750$
during sudden increases in MSD, as observed in
Fig.~\ref{fig:MSD_D_solid}(a).
The dynamics at $\nu=0.740$ and $\nu=0.750$ are fundamentally similar, with the primary difference being the frequency of particle activation (large displacements). At $\nu=0.740$, particle activations occur simultaneously and successively, resulting in a linear increase in MSD as shown in Fig.~\ref{fig:MSD_D_solid}(a). This behavior can be interpreted as a superposition of the sudden step-like increases observed at $\nu=0.750$.

To further investigate the microscopic diffusion mechanisms, we analyzed particle trajectories for
$N_{\mathrm{ev}}^*=300$
in regions exhibiting local diffusion, as shown in Fig.~\ref{fig:hopping}(bottom).  
The particle trajectories were colored according to the developmental stages with the number of 
events, as implemented in Refs.~\onlinecite{lam_2017, yip_2020}.
For displaying particle trajectories, we used the coarse-grained particle positions $\overline{\mathbf{r}_i}$ obtained by the following equation to mitigate thermal fluctuations in particle trajectories within the cage and observe the intrinsic motion,
\begin{equation}
\overline{\mathbf{r}_i}(t) = \frac{1}{\delta t}\int_{0}^{\delta t}
[\mathbf{r}_i(t+t') - \left\langle \Delta \mathbf{r}(t+t') \right\rangle] dt',
\label{eq:CG}
\end{equation}
where 
$\left\langle \Delta \mathbf{r}(t+t') \right\rangle$ is essentially the same as $\left\langle \Delta \mathbf{r} \right\rangle$ that used in Eqs.(\ref{eq:D}) and (\ref{eq:Dcpu}), and $\left\langle \Delta \mathbf{r}(t+t') \right\rangle=\sum_i[\mathbf{r}_i(t+t') - \mathbf{r}_i(0)]/N$.
In this study, we used the 
event
number 
$N_{\mathrm{ev}}^*$
instead of time $t$ to examine the dynamics of Monte Carlo simulations, setting $dt'=1\times N_{\mathrm{ev}}^*$
and $\delta t=100\times N_{\mathrm{ev}}^*$
in Eq.~(\ref{eq:CG}).

This analysis revealed motion reminiscent of the string-like hopping motion observed in
glassy systems\cite{miyagawa_1988, keys_2011, speck_2012, isobe_2016b, schoenholz_2016, yip_2020}. 
Such dynamic heterogeneity and hopping motions were considered intrinsic relaxation mechanisms in equilibrium at these highly packed states above the hard disk transition point, as these phenomena were observed across all methods employed in this study.

\begin{figure*}
    \centering
    \includegraphics[scale=0.14]{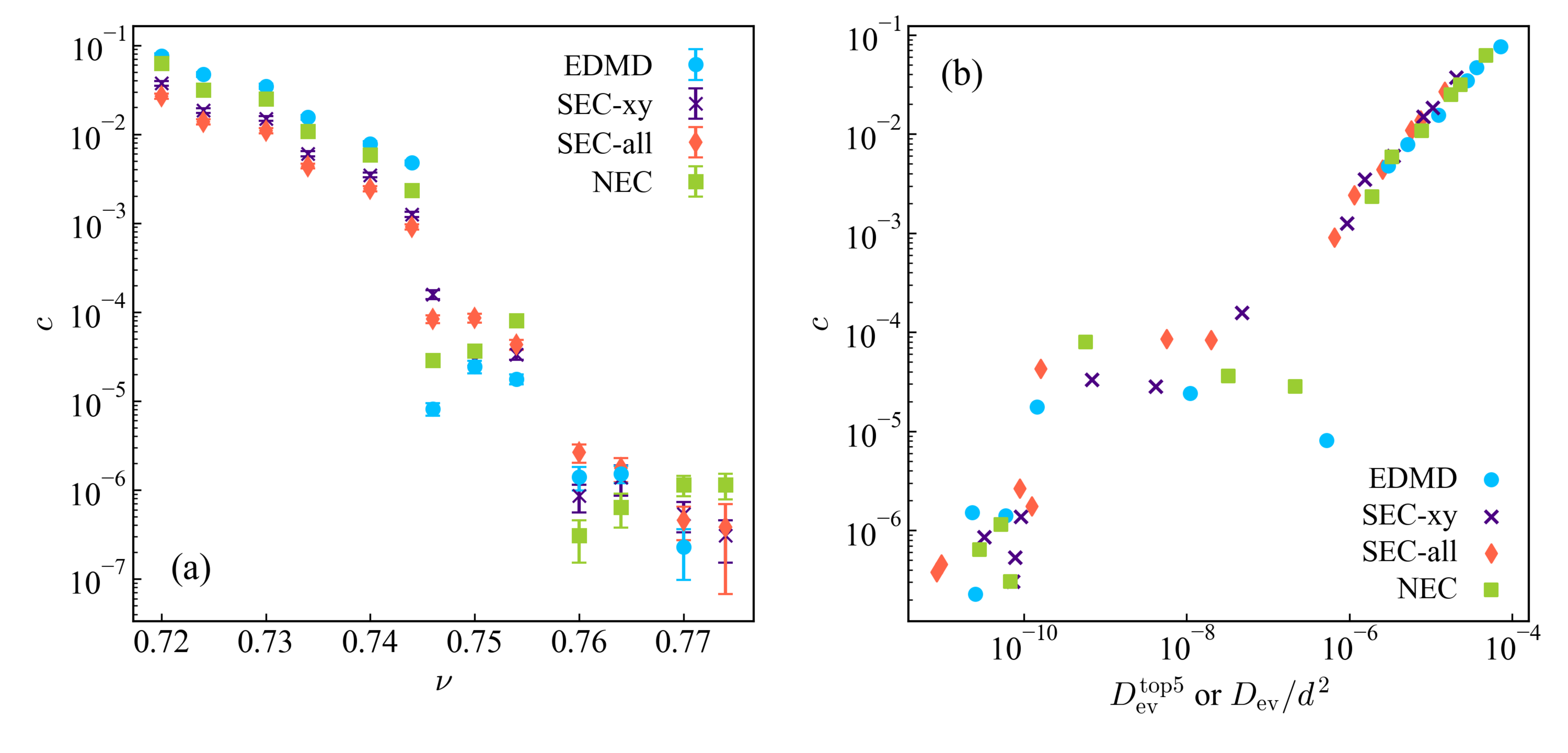}
    \caption{(a) Packing fraction dependence of the average hopping concentration $c$ in the solid phase for $N=65536$. 
In each ECMC variant, $L_{\mathrm{c}}^*$ or $T_{\mathrm{c}}^*=28$ was used. 
(b) Relationship between $c$ and the diffusion coefficients.}
    \label{fig:Dtop5_P}
\end{figure*}

To investigate the relationship between hopping motions and diffusion, we examined the correlation between the hopping concentration $c$ and the diffusion coefficient for each method. Following the methodology described in Refs.~\onlinecite{keys_2011, speck_2012, isobe_2016b}, $c$ is estimated using the following equation:
\begin{equation}
c = \left\langle \frac{1}{N}\sum^N_{i=1}\theta[ | \overline{\mathbf{r}_i}(N_{\mathrm{ev}}^*) - \overline{\mathbf{r}_i}(0) | - a] \right\rangle,
\end{equation}
where $\theta$ is the Heaviside step function.
$N_{\mathrm{ev}}^*$
and $a$ represent the characteristic number of 
events
and length of the hopping motion, respectively. In this study, these parameters were set to ($N_{\mathrm{ev}}^*$, $a/d$) = ($500$, $0.7$). The angle brackets denote an ensemble average over $200$ iso-configurations with different velocities for each of $5$ independent configurations, totaling $1000$ ensembles. To avoid detecting hopping caused by thermal vibrations, we also used the coarse-grained particle positions $\overline{\mathbf{r}_i}$ for the calculation of $c$.

Figure~\ref{fig:Dtop5_P}(a) illustrates the dependence of the average hopping concentration $c$ on the packing fraction $\nu$ for each method. For $\nu \le$
$0.744$, as the packing fraction increases, the values of $c$ decrease while maintaining the efficiency ranking among the methods, consistent with the diffusion efficiency order shown in Fig.~\ref{fig:MSD_D_solid}(c). Overall, the trend of $c$ closely parallels that of the 
event-based
diffusion coefficient. Indeed, as shown in Fig.~\ref{fig:Dtop5_P}(b), there is a strong positive correlation between $c$ and 
$D_{\mathrm{ev}}^{\mathrm{top5}}$
 (or $D_{\mathrm{ev}}$
 for EDMD). These findings suggest that in the solid phase, diffusion is predominantly governed by string-like hopping motions. Among the three ECMC variants used in this study, NEC exhibited the highest average hopping concentration at $\nu \le0.744$.

Note on evaluating and validating the range of $N_{\mathrm{ev}}^*$ for estimating diffusion coefficients in highly packed systems:
As $N_{\mathrm{ev}}$ is increased, a step-like behavior emerges. At $\nu = 0.750$, hopping motion occurs independently in time, resulting in step-like increases in MSD. We observed hopping motion at $N^*_{\mathrm{ev}} \le 10^6$ even for the shortest chain length in SEC-all, suggesting the minimal value for diffusion by hopping motion within shorter chain lengths or durations.
Regarding the upper limit of $10^7$, the typical waiting time for hopping motion (in this case, $N_{\mathrm{ev}}$) averaged over $25$ samples at $\nu = 0.750$ is estimated and found to be reasonable. 
For $\nu > 0.750$, it is important to note that applying the criterion for the range of $N_\mathrm{ev}^\ast$ used at $\nu \leq 0.750$ in the diffusive regime may result in inaccurate (artificial) values of the diffusion coefficient. This is because the waiting time for hopping events becomes longer, potentially leading to an underestimation of the diffusion coefficient. However, within
the parameter spaces we focus on, the calculated diffusion coefficients correlate nearly linearly with hopping concentration in the iso-configuration ensemble observed in Fig.~\ref{fig:Dtop5_P}(b), suggesting that this range is practically reasonable.

\section{\label{sec:4} Concluding Remarks}
In this paper, we systematically investigated the diffusional characteristics of ECMC variants (SEC-xy, SEC-all, and NEC) and EDMD in monodisperse hard disk systems for both liquid and solid phases. NEC often demonstrated superior efficiency in terms of diffusion coefficient per
event
and CPU time among ECMC variants, while SEC-xy showed better CPU-time efficiency in certain chain length regimes due to simplified implementation.

The diffusion coefficients exhibited a three-stage behavior with respect to chain length
or duration, attributed to increased collision events relative to simple accepted displacements (as in MCMC) and reducible sampling due to finite system size effects. Notably, NEC's diffusion coefficient remained independent of chain duration time in the limit of infinity and increased logarithmically with system size, suggesting potential divergence in the thermodynamic limit.

In
the system with $N = 65536$ in
the solid phase up to $\nu =0.744$, string-like hopping motions, which are often observed in the slow dynamics of supercooled liquids in glassy systems,
dominated diffusion, with NEC again showing the highest efficiency
in terms of event. However, diffusion efficiency generally decreased with increasing packing fraction, and diffusion coefficients vanished around $\nu = 0.760$.

These findings provide insights
into the optimization of slow equilibrium dynamics
in high-density and large-scale systems, especially in reducing practical computational costs.
In future work, we will focus on 
the system size dependency of diffusion dynamics and microscopic mechanism, not only in the equilibrium solid phase but also in the coexistence phase in equilibrium and/or during equilibration from non-equilibrium states to equilibrium. This includes investigating interference with phase transitions (previously studied by Ref.~\onlinecite{mugita_2021} using EDMD) using ECMC variants.
We will also explore the underlying mechanism of NEC's size-dependent efficiency, which reminds us of the logarithmic divergence of the diffusion coefficient due to vortex flow in the two-dimensional long-time tail problem\cite{alder_1967, alder_1969, alder_1970, isobe_2008, isobe_2009}.
These investigations have the potential to deepen
our understanding of optimal methods for  
efficiency in molecular simulations and would provide insights for developing optimal algorithms
in complex many-body systems.

\begin{acknowledgments}
The authors are grateful to Professor Werner Krauth, Professor Michael Engel, and Dr. Yoshihiko Nishikawa for stimulating and fruitful discussions. Special thanks are extended to Mr. Hirotaka Banno for investigating preliminary results, especially in the liquid phase for this project. Additionally, special thanks to Mr. Hiroaki Murase for providing the highly efficient code for EDMD. M.I. was supported by JSPS KAKENHI Grant Nos. 20K03785 and 23K03246. Part of the computations were performed using the facilities of the Supercomputer Center, ISSP, University of Tokyo. This research was conducted within the context of the International Research Project ``Non-Reversible Markov Chains, Implementations and Applications''.
\end{acknowledgments}

\section*{AUTHOR DECLARATIONS}

\subsection*{Conflict of Interest}
The authors have no conflicts to disclose.

\subsection*{Author Contributions}
{\bf Daigo Mugita}: Conceptualization (equal); Data Curation (lead); Formal Analysis (lead); Investigation (lead); Methodology (lead); Project Administration (equal); Software (lead);  Validation (lead); Visualization (lead); Writing - Original Draft (lead); Writing - Review and Editing (lead).
{\bf Masaharu Isobe}: Conceptualization (lead); Data Curation (supporting); Formal Analysis (supporting); Funding Acquisition (lead); Investigation (supporting); Methodology (equal); Project Administration (lead); Resources (lead); Software (equal); Supervision (lead); Validation (supporting); Visualization (supporting); Writing - Original Draft (lead); Writing - Review and Editing (lead).

\section*{DATA AVAILABILITY}
The data that support the findings of this study are available from the corresponding author upon reasonable request.

\bibliography{bibtex}

\end{document}